\documentclass[11pt]{article}    
    \usepackage[T1]{fontenc}
    \usepackage{mathpazo}

    \usepackage{graphicx}
    \makeatletter
    \def\maxwidth{\ifdim\Gin@nat@width>\linewidth\linewidth
    \else\Gin@nat@width\fi}
    \makeatother
    \let\Oldincludegraphics\includegraphics
    \renewcommand{\includegraphics}[1]{\Oldincludegraphics[width=.8\maxwidth]{#1}}
    \usepackage{caption}
    \DeclareCaptionLabelFormat{nolabel}{}
    \usepackage{adjustbox} 
    \usepackage{xcolor} 
    \usepackage{enumerate} 
    \usepackage{geometry} 
    \usepackage{amsmath} 
    \usepackage{amssymb} 
    \usepackage{textcomp} 
    \AtBeginDocument{%
    }
    \usepackage{upquote} 
    \usepackage{eurosym} 
    \usepackage[mathletters]{ucs} 
    \usepackage[utf8x]{inputenc} 
    \usepackage{fancyvrb} 
    \usepackage{grffile} 
    \usepackage{hyperref}
    \usepackage{longtable} 
    \usepackage{booktabs}  
    \usepackage[inline]{enumitem} 
    \usepackage[normalem]{ulem} 
    \title{On soil activation by cosmic rays at different altitudes}
    \author{
    	Yu.V. Stenkin$^{1,2}$ and O.B. Shchegolev$^1$  	
 \and
    	$^1$ Institute for Nuclear Research, Russian Academy of Science,  Moscow, Russia \\
    	$^2$ National Research Nuclear University MEPhI, Moscow, Russia
    } 
 
\begin{document}
   	
   	\label{firstpage}
    \maketitle

 \begin{abstract}

    Measuring radon-due neutron flux at various altitude (100, 1000, 1700, 4300 m above sea level) we found an evidence of significant increase of radon concentration  with altitude. It was also conirmed by direct radon measurements at high altitude. This allowed us to assume cosmic rays could take part in process of soil activation: they transform long-lived nuclei of uranium and thorium to nuclei with shorter life-time through specific nuclear reactions. If the resulting nuclei belong to the U-238 radioactive chain they can lead to producion of Ra-226 and then to Rn-222, thus significantly increasing its production at high altitudes where cosmic ray flux is high.
    
    keywords: thermal neutron; radon; cosmic rays
    
\end{abstract}

    \section{Introduction}\label{introduction}

    The theory on natural radioactivity postulates that radionuclides (for
example \(^{222}Rn\)) are in equilibrium with their parents in
radioactive decay chain. There are known four main radioactive chains,
starting from long-lived nuclides: U-238, Th-232, U-235 and Np-237. Rn-222 is product of U-238 decay chain, so it has to be in equilibrium with it's long-lived parents. It means that quantity of this nuclide depends only on quantity of U-238 randomly spread over the Earth crust. While measuring natural thermal neutron flux at sea level we noticed that it's barometric
coefficient is \textasciitilde{} 0.8-0.9\%/mm Hg. On one hand we know that
barometric coefficient of cosmic ray hadrons is \textasciitilde{} 1\%/mm
Hg. On another hand we assume that there are two main natural neutron
flux sources: cosmic ray hadrons (producing evaporation neutrons in
collisions with nuclei of soil and air, those are thermalized later) and
natural radioactivity (\(\alpha\)-decay nuclei, resulting in 
neutrons production in \((\alpha,n)\) reaction). Taking above into account we can
suppose that natural neutron flux consists by \textasciitilde{}80-90\% of
neutrons produced by cosmic rays and 10-20\% by natural radioactivity
{[}1{]}. We have tested this assumption measuring thermal neutron
flux underground: at a level of \textasciitilde{} 2.5 m of soil,
barometric coefficient is \textasciitilde{} 0.6-0.7 \%/mm Hg. Later we
have measured thermal neutron flux at few different altitudes: 1000 m
(Grand Sasso, Italy), 1700 m (Baksan, Russia), 3600 m (Lhasa, Tibet,
China), 4300 m (Yangbajing, Tibet, China). We supposed that a fraction of
neutrons produced by cosmic ray hadrons has to grow up with altitude because cosmic ray flux rises with altitude while natural radioactivity supposed to have no such dependence. But as we have found: barometric coefficient does not depend on altitude [1]. The reason of such behavior could be found if one supposes that quantity of natural radioactive nuclides producing neutrons is also in equilibrium with integral cosmic ray flux accumulated for a long periods (thousands years). In other words:  quantity of radioactive nuclides in upper layer of soil is also proportional to cosmic ray flux and should increase with altitude. It has been confirmed by radon meters data obtained at Yangbajing (4300 m). Concentration of radon and thus quantity of \(\alpha\)-active nuclides there is  \textasciitilde{} 8 times higher than in Moscow or in Beijing. Let's try to explain it.

    \section{ Hypothesis}\label{hypothesis}

    Our hypothesis is really simple. Let us consider two the most rife radioactive chains.  U-238 has a half-life \(t = 4.47*10^9 years\) and it is the second of most common heavy
radionuclide in Earth crust. Th-232 has half-life \(t = 1.41*10^{10} years\) and it is the first one. Rn-222 is a product of U-238 decay chain and in common view it is in equilibrium with U-238.
Now let us look to reactions \[^{232}Th (n,3n) ^{230}Th\] and \[^{232}Th (\gamma,2n) ^{230}Th\] The first one has cross section \(\sigma \approx 0.85\pm0.15\) barns {[}2{]} and the second one -
\(\sigma=0.3\pm0.05\) barns {[}3,4,5{]} for neutrons and \(\gamma\) with
\(E\approx(14-15)MeV\). These reactions result in production of Th-230 belonging 
to the U-238 decay chain and having half-life t = 7.7*10\^{}4, i.e. by a factor of  6*10\^{}4 higher than U-238 has. Therefore, the production rate of Ra-226 being both product of Th-230 decay and parent nuclide for Rn-222, would be much higher. On the other hand, neutrons and gammas of energy \textasciitilde{}(10-20) MeV are produced by cosmic rays in upper layer of soil (4-5 m). It means that concentration of Th-230, Ra-226, Rn-222 (and its products) in upper soil layer and thus flux of radon-due neutrons,  should increase with long-term cosmic ray integral flux, while immediate neutron flux is proportional to immediate cosmic ray flux. It is obvious that ratio of  immediate to accumulated cosmic ray fluxes does not depend on altitude in the fist approximation.  

    \section{ Simulation}\label{simulation}

    Using modern GEANT4.10 package code we have simulated soil containing
0.0013\% of Th-232 (it corresponds to average natural quantity)
bombarded with neutrons of energy 20 MeV. We searched for number of
produced Th-230 nuclei. As a preliminary result we obtained
\(n\approx (2\pm1.4)*10^{-6}\) nuclei of Th-230 per 1 neutron of 20 MeV.
Therefore, these reactions do exist and we have to simulate these processes in more details.

    \section{Exerimental results}\label{exerimental-results}

    Averaged counting rate of radon meters used by ARGO collaboration
{[}6{]} in YangBaJing guest house rooms and open experimental hall was
\textasciitilde{} 400 \(Bq/m^3\). It is much higher than in buildings at
the sea level in Europe (for example Netherlands) \textasciitilde{} 40
\(Bq/m^3\). Also radon activity in Switzerland in Alps is
\textasciitilde{} 200 Bq/m\^{}3 {[}7{]}. By our measurements counting
rate of neutron detectors is \textasciitilde{} 8 times higher in YBJ
than that at sea level {[}8{]}. It means that at 4300 m a.s.l. concentration of Th-230 nuclei is approximately 8 times higher than at sea level.

    \section{Conclusion}\label{conclusion}

    We suggest here a way to explain very high Rn concentration at high altitudes.  Cosmic ray neutrons and gamma-quanta's produce Th-230 from soil Th-232 due to nuclear reactions and thus increase radioactivity of upper soil level  and accelerate natural decay chain process by a factor of $\sim$\(10^5\). Such a large factor can compensate low probability of the above reactions and can result in significant acceleration of the Th-230 production rate.
This reactions replace radioactive nuclei of Th-232 family with nuclei of U-238 chain family resulting in  Rn-222 production. It leads to significant increase of Rn-222 concentration in high mountains due to exponential increase of the cosmic rays flux with altitude. This hypothesis is confirmed by our preliminary GEANT4.10 simulations and experimental measurements of radon flux and thermal neutron flux at different altitudes.

\section{Acknowledgements}

 The work was supported by the Russian Foundation for Basic Research, project Nos. 16-32-00054, and 16-29-13067 and by "Fundamental properties of matter and astrophysics" Program of the Presidium of the Russian Academy of Sciences.


\begin{thebibliography}{20}
	
	\bibitem{1}	Alekseenko, V.V., Gromushkin, D.M. and Sten'kin. Comparative
	measurements of thermal neutron fluxes at ground level at the baksan
	neutrino observatory and LGNS laboratory. Y.V. Bull. Russ. Acad. Sci.
	Phys. Vol. 75, p. 857 (2011).
	\bibitem{2}	M.H.Mctaggart and H.Goodfellow. Measurements at 14 MeV neutron energy of
	the (n,2n) cross section of beryllium and the (n,3n) cross section of
	thorium.J.Nuclear Energy A and B (Reactor Sci. and Technol.), Vol.17, p.437
	(1963).
	\bibitem{3}		 V.V.Varlamov and N.N.Peskov. Evaluation of (gamma,xn), (gamma,sn),
	(gamma,n), (gamma,2n), and (gamma,f) reactions cross sections for
	actinides nuclei 232Th, 238U, 237Np, and 239Pu: consistency between data
	obtained using quasimonoenergetic annihilation and bremsstrahlung
	photons. Moscow State Univ. Inst.of Nucl. Phys.Reports, No.2007, p.8/829
	(2007).
	\bibitem{4}	A.Veyssiere, H.Beil, R.Bergere, P.Carlos, A.Lepretre and K.Kernbach. A
	Study of the Photofission and Photoneutron Processes in the Giant Dipole
	Resonance of 232Th, 238U and 237Np. Nuclear Physics, Section A, Vol.199,
	p.45 (1973).
	\bibitem{5}		A.Veyssiere, H.Beil, R.Bergere, P.Carlos, A.Lepretre and K.Kernbach. A
	Study of the Photofission and Photoneutron Processes in the Giant Dipole
	Resonance of 232Th, 238U and 237Np. Nuclear Physics, Section A, Vol.199,
	p.45 (1973).
	\bibitem{6}	B. Bartoli et al. Radon contribution to single particle counts of
	the ARGO-YBJ detector. Radiation Measurements, Vol.68 p.42-48 (2014).
	\bibitem{7}	United Nations Environment Programme. Radiation: Doses, Effects,
	Risks. B. Blackwell, 88(1991).
	\bibitem{8} Yuri Stenkin, Victor Alekseenko, et al. Seasonal and Lunar Month Periods Observed in
	Natural Neutron Flux at High Altitude. Pure and Appl. Geophys. 174 (2017), 2763–2771.

\end{thebibliography}
    \end{document}